# The Spillover Effects of Peer AI Rinsing on Corporate Green Innovation

Li Wenxiu[1] Wen Zhanjie[1] Xia Jiechang[2] Guo Jingqiao[3]

[1] School of Economics and Trade, Guangdong University of Finance, Guangzhou, Guangdong, 510521

[2] Institute of Finance and Economics, Chinese Academy of Social Sciences, Beijing, 100006

[3] Department of Computer Science, Faculty of Science, Hong Kong Baptist University, Hong Kong, 999077

(School of Economics and Trade, Guangdong University of Finance, Guangzhou, Guangdong 510521)

**Abstract：** At a time when the phenomenon of 'AI washing' is quietly spreading, an increasing number of enterprises are using the label of artificial intelligence merely as a cosmetic embellishment in their annual reports, rather than as a genuine engine driving transformation. A test regarding the essence of innovation and the authenticity of information disclosure has arrived. This paper employs large language models to conduct semantic analysis on the text of annual reports from Chinese A-share listed companies from 2006 to 2024, systematically examining the impact of corporate AI washing behaviour on their green innovation.The research reveals that corporate AI washing exerts a significant crowding-out effect on green innovation, with this negative relationship transmitted through dual channels in both product and capital markets. Furthermore, this crowding-out effect exhibits heterogeneity across firms and industries, with private enterprises, small and medium-sized enterprises (SMEs), and firms in highly competitive sectors suffering more severe negative impacts from AI washing. Simulation results indicate that a

General Project of the National Social Science Fund for the Year 2025: Research on Theoretical Mechanisms and Path Selection for Enhancing the Resilience of Industrial and Supply Chains through the Integration of Data and Reality (Project No. 25BJL034); General Project of Guangdong Provincial Social Science Planning 2025 "Evaluation of the Effect, Impact Mechanism, and Implementation Path of Integrating Data and Reality to Empower Supply Chain Resilience Enhancement" (Project No. GD25CYJ14); Guangzhou's 14th Five Year Plan Project "Multidimensional Effects and Collaborative Governance Path of Artificial Intelligence on Regional Employment Market Structure" (Project No. 2025GZGJ118)

[Author Introduction] Li Wenxiu (1978-), female, from Hanchuan, Hubei, is a researcher at Guangdong University of Finance and holds a PhD in Economics. Research direction: Industrial Economics and International Trade, email: lwxasu@126.com. Wen Zhanjie (1992-), male (Han), from Lufeng, Guangdong, is a lecturer at Guangdong University of Finance. Research direction: Causal inference and machine learning, email: 70-154@gduf.edu.cn The corresponding author. Xia Jiechang (1964-), male, is a researcher and doctoral supervisor at the Institute of Finance and Economics Strategy, Chinese Academy of Social Sciences. My main research directions are digital economy, service economy, and consumer economy. Email: xiajch@cass.org.cn Guo Jingqiao (2000-), female (Han), from Yinchuan, Ningxia, holds a Master of Science in Information Technology Management from Hong Kong Baptist University.Research direction: Causal inference and machine learning. Email: guojingqiao666@gmail.com .

combination of policy tools can effectively improve market equilibrium.Based on this, this paper proposes that the government should design targeted support tools to 'enhance market returns and alleviate financing constraints', adopt a differentiated regulatory strategy, and establish a disclosure mechanism combining 'professional identification and reputational sanctions' to curb such peer AI washing behaviour.

**Keywords**: AI washing; green innovation; information disclosure; large language models; agent-based modelling

## I. Introduction

Currently, the global economy is undergoing a dual transformation characterised by profound digitalisation and a shift towards a green, low-carbon economy. Artificial intelligence (AI) technology is regarded as the core engine of the Fourth Industrial Revolution (Babina et al., 2024). China has successively issued documents such as the 'Development Plan for the New Generation of Artificial Intelligence' and the '14th Five-Year Plan for the Development of the Digital Economy', explicitly stating its intention to secure a leading position in global AI technology.Concurrently, the introduction of the 'dual carbon' goals signifies a fundamental shift in China's economic development model, with green technological innovation becoming the key pathway to achieving carbon neutrality (Du et al., 2015). However, in an environment of information asymmetry, the capital market's frenzied pursuit of AI concepts has spawned a systemic risk: a severe disconnect between actual applications and disclosed content, giving rise to the phenomenon of 'AI washing' (Li et al., 2025).More notably, this behaviour exhibits contagious spread within the industry; the proliferation of peer-driven AI washing not only exacerbates market information asymmetry (Akerlof, 1970) but may also intensify through competitive effects and signal dilution (Spence, 1973). Although genuine digital technology applications can promote green innovation by enhancing resource allocation efficiency and optimising production processes (Jin Xingye et al., 2024),However, according to Akerlof's (1970) 'lemons' market theory, when the market is flooded with false information, the mechanism of adverse selection prevents firms that genuinely invest resources in technological innovation from being adequately compensated for their costs, ultimately driving out high-quality firms and leading to market contraction. Furthermore, greenwashing by firms can also exert a crowding-out effect on green innovation by peers through mechanisms in both product and capital markets (Cai Zhen and Wan Zhao, 2024).Consequently, when AI greenwashing becomes a widespread phenomenon within the industry, digitalisation strategies and green development strategies may compete for resources rather than complementing one another.Firms will direct their limited R&D budgets and managerial attention towards AI-driven conceptual packaging rather than green technological innovation (Guo Junjie et al., 2024); the effectiveness of government incentives for green innovation will be diluted; and financial institutions will struggle to identify firms with genuine green technological capabilities (Tang Song et al., 2020), resulting in a systemic crowding-out effect on substantive innovation—particularly green innovation—across the entire industry. When the industry as a whole becomes mired in conceptual hype rather than genuine technological investment, the market recognition required for corporate green innovation declines, the financing environment deteriorates, and strategic resources become fragmented, ultimately leading to execution deviations between the two strategies at the micro level.

As the world's largest market for AI applications and a key driver of green technological innovation, China must pay particular attention to the damage caused to the innovation ecosystem by the deterioration of information disclosure quality at the industry level. Especially during the critical phase of manufacturing transformation and upgrading, as well as the cultivation of strategic emerging industries, widespread AI 'greenwashing' within the sector may lead to a 'bad money drives out good money' scenario, where enterprises genuinely investing in green technology R&D lose their innovative drive because the market fails to recognise their true value (Delmas & Burbano, 2011). However, existing research lacks a systematic analysis of the negative spillover effects of distorted information disclosure at the industry level. In particular, the mechanism by which peer AI greenwashing in the digital technology sector influences corporate green innovation decisions through market competition and signalling mechanisms has not yet been fully elucidated (Zhang Detao et al., 2024). In fact, the disruption of market order caused by distorted information disclosure has already aroused high levels of vigilance within the international community. Both the EU's Sustainable Finance Disclosure Regulation and the US Securities and Exchange Commission's enforcement actions against 'greenwashing' reflect the importance attached to the systemic risks of false disclosures (Lyon & Maxwell, 2011). Against this backdrop, this study focuses on the crowding-out effect of peer AI greenwashing on corporate green innovation, aiming to address three core questions: Does the AI greenwashing behaviour of other firms within the industry significantly inhibit a firm's own green innovation? Through which market mechanisms is this negative spillover effect transmitted? Do the degrees of impact vary across different competitive environments and firm characteristics?

This study extends greenwashing research to the field of AI washing, revealing a new form of information distortion in corporate disclosures within the context of digital technology, thereby enriching the application of information asymmetry theory in innovation management; the proposed dual crowding-out mechanism integrates perspectives from both product and capital markets, deepening our understanding of the complexity of corporate innovation decision-making.Furthermore, the study innovatively employs large language models for text semantic analysis, overcoming the limitations of traditional keyword matching methods and providing a new paradigm for measuring unstructured data. It also introduces agent-based modelling and simulation methods, addressing the shortcomings of static empirical analysis in terms of dynamic evolution and policy simulation, thereby achieving a methodological breakthrough in quantitative research (Guo Junjie et al., 2024).

## II. Measurement of AIwashing and its Real-World Depiction

### (1) Measurement Methods for AIwashing

#### 1. Theoretical Framework

Existing literature primarily employs three methods to measure information manipulation: sentiment dictionary-based information whitewashing measurement, discourse-discrepancy-based disconnection measurement, and third-party evaluation-based identification (Cai Zhen and Wan Zhao, 2024).The sentiment dictionary method relies on a predefined vocabulary to judge the bias of information; however, the completeness and applicability of such dictionaries in a Chinese context remain to be verified, and it is difficult to distinguish the actual meaning of keywords within specific contexts. The discrepancy method identifies greenwashing by comparing the gap between corporate disclosures and actual actions; however, such studies are often limited to companies that publish specialised disclosure reports, resulting in restricted sample coverage.Although the third-party evaluation method possesses objectivity, it struggles to quantify the extent of greenwashing and suffers from a small sample size. In light of this, this study draws upon the methodological framework for constructing greenwashing variables developed by Cai Zhen and Wan Zhao (2024), combined with Li et al.’s (2025) approach to measuring AIwashing, and employs large language model text analysis to construct an AIwashing variable.This method enables the identification of the actual meaning of AI keywords within specific contexts, thereby avoiding the inclusion of irrelevant statements in the analysis. Furthermore, it effectively distinguishes between substantive and descriptive AI statements, accurately capturing the extent of the disconnect between corporate AI disclosures and actual practices.Jin Xingye et al. (2024) successfully utilised large language models to measure corporate digital transformation, validating the effectiveness of this method. Furthermore, the method is suitable for large-sample, long-time-series studies, covering all listed companies rather than being limited to specific industries or time periods, thereby ensuring the generalisability and comparability of the research.

## 2. Variable Construction

The construction of the AIwashing variable follows an analytical logic comprising text extraction, statement classification, corporate indicators, and industry indicators. This study selects the 'Management Discussion and Analysis' section of listed companies' annual reports as the text source. This serves as the core channel through which management communicates corporate strategy, operational performance, and future plans to investors; as information disclosure is relatively unrestricted, it effectively reflects the true characteristics of corporate AI disclosure behaviour.This selection draws on the analytical approach employed by Li et al. (2025) in their analysis of earnings conference call transcripts, focusing on scenarios where management proactively conveys information with a high degree of freedom. The construction of the AI keyword corpus comprehensively considers the evolutionary trajectory of AI technology and the reporting conventions of Chinese enterprises. Core terms include artificial intelligence, machine learning, deep learning, neural networks, natural language processing, computer vision, intelligentisation, algorithms, data mining, smart manufacturing, industrial internet, and intelligent decision-making.In light of the breakthrough developments in generative AI technology, the lexicon also incorporates emerging terms such as generative AI, large language models, ChatGPT, Wenxin Yiyan, Transformer, pre-trained models, multimodal, AIGC, and computing infrastructure. Python was used to extract complete sentences containing the aforementioned keywords from management discussion and analysis texts, thereby constituting the corpus for classification.

During the sentence classification stage, this study employs Baidu's ERNIE large language model to perform a three-way classification of AI-related sentences, building upon the core concept of distinguishing between 'AI talk' and 'AI walk' proposed by Li et al. (2025). The classification criteria comprise three categories. The first category consists of non-AI sentences, coded as 0, where the meaning of the AI keywords in that context is unrelated to AI technology.The second category comprises substantive AI statements, coded as 2, which contain objective statements regarding quantifiable AI inputs, outputs, and outcomes, such as the number of AI patent applications filed by the company increasing to 50 this year or AI product revenue reaching 50 million yuan. The third category consists of descriptive AI statements, coded as 1, referring to statements regarding AI plans, visions, and philosophies that are difficult to verify, such as the company's commitment to vigorously developing AI technology or its aspiration to become a leading enterprise in the AI sector.The model was trained using a training set constructed from approximately 6,000 manually annotated sample sentences, ensuring a balanced distribution of the three statement categories and accurate annotation. Fine-tuning was performed using the ERNIE model, achieving a classification accuracy of over 85% on an independent validation set. The trained model was subsequently applied to the entire corpus to enable automated classification.

The calculation of a company’s AIwashing level follows the greenwashing measurement method proposed by Cai Zhen and Wan Zhao (2024), using the number of descriptive AI statements to quantify a company’s AIwashing tendency. Specifically, the formula for the AI_Tone indicator is:

$$AI_Tone_{i,\ t} = \frac{N_{描述性AI语句,\ i,\ t}}{N_{总AI语句,\ i,\ t}}$$

where $N_{描述性AI语句,\ i,\ t}$ denotes the number of descriptive AI statements in company i’s annual report for year t, and $N_{总AI语句,\ i,\ t}$ denotes the total number of AI-related statements for that company in the same year. A higher AI_Tone value indicates a greater proportion of descriptive content in the company’s AI disclosures, relatively less substantive content, and a stronger tendency towards AIwashing.

To further characterise the absolute level of corporate AIwashing, this paper constructs a firm-level AIwashing index. Drawing on the research approach of Shen et al. (2023), the following formula is used for calculation:

$$AIwashing_{i,\ t} = \ln\left(1 + AI_Tone_{i,\ t} \times N_{总AI语句,\ i,\ t}\right)$$

This indicator takes into account both the intensity and scale of a firm’s AIwashing; the logarithmic transformation helps mitigate the impact of outliers, resulting in a more reasonable distribution of the indicator.

Taking into account the externalities of peer firms’ behaviour, this paper further constructs an industry-level peer AIwashing index. Drawing on the measurement method for peer greenwashing proposed by Cai Zhen and Wan Zhao (2024), the calculation formula is:

$$AIwashing_peer_{j,\ t} = \frac{1}{N_{j,\ t} - 1} \sum_{i \in j,\ i \neq focal} AIwashing_{i,\ t}$$

where $AIwashing_peer_{j,\ t}$ denotes the degree of peer AIwashing in industry j in year t, and $N_{j,\ t}$ represents the number of firms in that industry in the same year; the summation excludes the target firm itself, calculating only the average level of AIwashing among peer firms. This indicator reflects the overall AIwashing climate within the industry in which a firm operates and serves as the core explanatory variable for studying the crowding-out effect of peer AIwashing on a firm’s green innovation.

## 3. Empirical Findings on Peer AIwashing and Green Innovation

To illustrate the relationship between peer AIwashing and green innovation, this study conducted a preliminary correlation analysis. Based on a sample of Chinese

A-share listed companies from 2006 to 2024, the Pearson correlation coefficient between the peer AIwashing index and a firm's green patent output was calculated to be -0.1817, which is significant at the 1% level. This negative correlation preliminarily supports the research hypothesis of this paper, namely that the higher the degree of peer AIwashing, the lower a firm's green innovation output.

Further analysis was conducted using bivariate regression, with the logarithm of the number of green patents granted as the dependent variable and the peer AIwashing index as the independent variable, whilst controlling for year and industry fixed effects.The regression results indicate that the coefficient for peer AIwashing is -0.2391, with a t-value of -6.82, which is significantly negative at the 1% level. This implies that for every one-unit increase in the peer AIwashing index, a firm's green patent output will decrease by approximately 23.91%. This empirical finding provides preliminary evidence for subsequent theoretical analysis and empirical testing, and highlights the importance of studying the crowding-out effect of peer AIwashing on green innovation.

## III. Mechanism Analysis and Research Hypotheses

### 1. The higher the degree of peer AIwashing, the lower the firm's green innovation output

According to the 'lemons market' theory (Akerlof, 1970) and signalling theory (Spence, 1973), when some firms in an industry engage in AIwashing, they send false signals of 'technological superiority' to the market. This makes it difficult for investors and consumers to distinguish firms' genuine technological capabilities, thereby exacerbating information asymmetry at the industry level.In this environment, the genuine value signals transmitted by firms committed to substantive green innovation are drowned out by noise; their innovative efforts struggle to gain full market recognition and returns, ultimately leading to a decline in innovation incentives (Zhou et al., 2023).Consequently, the higher the degree of AIwashing among peers, the lower a firm's green innovation output. AIwashing by peer firms exerts a crowding-out effect on a firm's green innovation through dual mechanisms in product and capital markets, leading to a significant reduction in the number of green patents held by the firm. Based on this, the following hypothesis is proposed:

**H1: The extent of AIwashing among peers is significantly negatively correlated with a firm's green innovation output.**

### 2. The Mechanisms of Market Share Erosion and Profit Crowding Out in Peer AIwashing

Peer AI washing exerts a crowding-out effect on firms' green innovation by weakening their market power, increasing competitive costs, and reducing profitability. Specifically, when peers exaggerate their AI applications, they may temporarily attract consumers or secure orders through misleading publicity, thereby squeezing the market share of firms that do not engage in AI washing (Bloom & Van Reenen, 2010).To counter this competition, firms may be forced to increase non-productive expenditures, such as marketing, to clarify or counteract false information (Gurun et al., 2022), resulting in resources being diverted away from long-term green technology R&D. Simultaneously, declining market share and rising competitive costs directly undermine a firm's profitability (Li et al., 2023), reducing the internal cash flow available for high-risk green innovation.Consequently, the higher the degree of peer AIwashing, the smaller the firm's market share, the higher its sales expense ratio, and the weaker its profitability, resulting in lower green innovation output. Based on this, the following hypothesis is proposed:

**H2: Peer AIwashing indirectly inhibits a firm's green innovation by reducing its market share and profitability.**

**3. Mechanisms of Resource Crowding and Exacerbated Financing Constraints Due to Peer AIwashing**

According to resource dependence theory and credit rationing theory (Stiglitz & Weiss, 1981), when widespread technological information distortion occurs within an industry, external investors and financial institutions may adopt a 'one-size-fits-all' prudential strategy to mitigate risk, thereby tightening credit supply to the entire industry or firms of the same type (Cheng et al., 2024).Concurrently, limited government R&D subsidies and innovation funds may be captured by firms adept at "packaging" rather than substantive action (Howell, 2017), leading to resource misallocation. This leaves firms genuinely engaged in green innovation facing more severe difficulties in debt financing and reduced policy-driven financial support—that is, intensified financing constraints—which in turn curtails their long-term R&D investment (Brown et al., 2012).Consequently, peer AIwashing inhibits green innovation by crowding out financing resources and government subsidies, thereby exacerbating firms' financing constraints. Specifically, the higher the degree of peer AIwashing, the weaker a firm's debt financing capacity, the fewer government subsidies it receives, the more severe its financing constraints, and consequently, the lower its green innovation output. Based on this, the following hypothesis is proposed:

**H3: Peer AIwashing indirectly inhibits green innovation by intensifying the crowding out of financing resources and strengthening firms' financing constraints.**

**4. The crowding-out effect of peer AIwashing on green innovation varies across different enterprises and market environments**

From the perspective of market power and risk buffers, firms with higher market share, stronger commitments to social responsibility and lower financial risk possess greater resilience to market fluctuations and typically place greater emphasis on long-term reputation (Flammer, 2015). When peers engage in AIwashing, if such firms persist with substantive innovation, they may face greater relative competitive disadvantages and suffer greater losses from their signals being obscured; consequently, their innovation decisions may be more sensitive to peers' AIwashing behaviour, resulting in a more pronounced crowding-out effect. From the perspective of adverse selection risks and supervisory governance, in market environments characterised by high information transparency and strong external oversight—such as those with high institutional investor ownership (Bushee, 1998) and dense analyst coverage (He & Tian, 2013)—the market's information processing capacity is stronger, and mechanisms for identifying and penalising 'pseudo-innovation' are more effective. This can partially alleviate the overall information asymmetry caused by peer AI-washing, ensuring that the value of genuinely innovative firms is recognised, thereby mitigating the negative spillover effects of peer AI-washing. Consequently, the crowding-out effect of peer AI-washing on green innovation varies across different firms and market environments. Based on this, the following hypothesis is proposed:

**H4a: The stronger a firm's market power, the better its social responsibility performance, and the more financially sound it is, the stronger the crowding-out effect of peer AI-washing on its green innovation.**

**H4b: The higher the transparency of the information environment in which a firm operates, the weaker the crowding-out effect of peer AI-washing on its green innovation.**

## IV. Model Specification and Variable Explanation for the ' '

### (1) Model Specification

A two-way fixed-effects model effectively mitigates omitted variable bias by eliminating unobservable firm-specific differences through firm-level fixed effects and controlling for the influence of common time trends via time-series fixed effects. This study employs a two-way fixed-effects model to examine the impact of corporate AI washing on green innovation, with the baseline regression equation set as Equation (1). Robust adjustments have been made to standard errors in the regression, and clustering has been applied at the firm level to address potential issues of heteroscedasticity and serial correlation (Cai Zhen and Wan Zhao, 2024).

$$gin_{it} = \alpha_0 + \alpha_1 AIwashing_{it} + \sum_k \gamma_k Controls_{it}^k + \mu_i + \lambda_t + \varepsilon_{it}$$

Here, the subscript i denotes the firm, and t denotes the year. The dependent variable $gin_{it}$ represents firm i's level of green innovation in year t, whilst the core independent variable $AIwashing_{it}$ represents firm i's degree of AI washing in year t. $Controls_{it}^k$ represents a set of control variables, including firm size, financial characteristics, and governance structure. $\mu_i$ is the firm fixed effect, controlling for firm-specific heterogeneity that does not vary over time, such as corporate culture and management style. $\lambda_t$ is the time fixed effect, absorbing time-series shocks such as macroeconomic fluctuations and changes in the policy environment. $\varepsilon_{it}$ is the random error term, assumed to satisfy classical assumptions. The key coefficient $\alpha_1$ measures the net effect of a firm's AI adoption on green innovation; if hypothesis H1 holds, $\alpha_1$ is expected to be significantly negative.

### (2) Variable Descriptions and Data Sources

#### 1. Dependent Variable

Green innovation is the core dependent variable in this study, measured by the number of green patent applications filed by firms. Green patents directly reflect a firm's innovation inputs and outputs in the field of environmental protection technologies and, compared to traditional innovation indicators, better reflect a firm's substantive contribution to environmental sustainability (Cai Zhen and Wan Zhao, 2025). The study follows the IPC Green List published by the World Intellectual Property Organisation to identify patent technology categories related to environmental protection, resource conservation, and pollution control. Given the right-skewed distribution of patent counts, the raw data is log-transformed; specifically, the natural logarithm of the number of green patent applications plus one is taken, denoted as gin. This treatment mitigates heteroscedasticity whilst retaining observational data for firms with zero patents, thereby enhancing the robustness of the regression results. Existing research also commonly employs the number of green patent applications as a proxy for green innovation, providing methodological support for this study (Shen Yi, 2023; Jin Xingye et al., 2024).

#### 2. Core Explanatory Variable

The level of corporate AI washing (Aiwashing) serves as the core explanatory variable in this study. First, the Wenxin model is utilised to classify AI-related statements in the 'Management Discussion and Analysis' section of annual reports, distinguishing between descriptive statements and substantive applications. Descriptive statements merely mention concepts at a theoretical level, lacking specific technical details and application scenarios, whereas substantive applications include clear technical pathways, application outcomes, or input-output information.By calculating the proportion of descriptive statements, we construct an AI tone indicator (AI_Tone), which is then weighted against the total number of AI-related terms (N_total). Finally, after logarithmic transformation, we obtain the firm-level AI washing index.

The peer AI washing level (AIwashing_peer) serves as an auxiliary explanatory variable, reflecting the average degree of AI washing among other firms within the industry. This variable is constructed using the mean of AI washing indices from other firms in the same industry and year, thereby capturing the industry-level climate of information disclosure and competitive pressure. When AI washing is prevalent within an industry, the difficulty of market identification for individual firms increases, and the problem of information asymmetry becomes more severe.The introduction of the peer AI-washing level helps distinguish the interaction between a firm's own behaviour and the industry environment, providing an important analytical perspective for mechanism testing (Li et al., 2025).

**3. Control Variables**

This study incorporates a range of firm characteristics and governance variables as control variables to mitigate the influence of confounding factors. Firm size (size) is measured using the natural logarithm of total assets, as larger firms typically possess greater resources to support innovation activities. The debt-to-asset ratio (lev) reflects a firm's level of financial leverage; high debt may constrain a firm's capacity for innovation investment. Return on assets (roa) measures a firm's profitability, with sound financial performance providing the financial foundation for green innovation.Firm age (age) is measured using the logarithm of the number of years since listing; mature firms may exhibit more stable innovation path dependencies. Governance structure variables include equity concentration, board size, and the proportion of independent directors. The shareholding ratio of the largest shareholder (top1) reflects the degree of equity concentration; moderate concentration is conducive to fostering long-term innovation incentives. Board size is measured using the logarithm of the total number of directors (board); the supervisory function of the board may influence a firm's innovation decisions.The proportion of external independent directors (indep) can provide professional advice and constrain short-sighted behaviour by management. Furthermore, the study controls for industry fixed effects and year fixed effects, accounting for industry characteristics that do not change over time and the impact of macroeconomic cycles, thereby making the identification strategy more rigorous (Guo Junjie et al., 2024).

(3) Data Sources

This study selects Chinese A-share listed companies from 2006 to 2024 as the research sample, constructing a comprehensive analytical framework through the integration of multi-source data. Sample selection adheres to the following criteria to ensure the reliability and validity of the research conclusions.Financial and insurance firms were excluded from the study, as their business models and financial characteristics differ significantly from those of real economy enterprises. Companies subject to special treatment (ST and PT) were also removed, as these firms often face delisting risks and may exhibit abnormal business behaviour. Observations with missing values for key variables were prudently excluded to prevent data quality issues from affecting the accuracy of the empirical results.Following systematic screening, a total of 55,745 annual observations from 5,424 listed companies were obtained, forming a large-scale, long-span unbalanced panel dataset.

The research data was derived from the integration of multiple authoritative databases. The construction of corporate AI screening indicators was based on annual report text information provided by the Guotai-An database; the Wenxin large language model was employed to conduct intelligent analysis of the Management Discussion and Analysis section, thereby identifying the degree of authenticity of AI-related statements made by companies.Data on green innovation outputs were sourced from the patent database of the China National Intellectual Property Administration, with green patent applications and grants identified using the Green List of the International Patent Classification (IPC). Corporate financial data and governance characteristics were obtained from the Guotai-An and Wind databases, including fundamental variables such as asset size, profitability and capital structure. Macroeconomic data and industry-specific variables were sourced from the National Bureau of Statistics and the China Industrial Enterprise Database, providing the necessary control variables for the study.

## IV. Analysis of Empirical Results

### (1) Descriptive Statistics

The descriptive statistics of the research sample reveal the basic distribution characteristics and data quality of the variables (see Table 1). The mean value of the green innovation variable gin is 0.8426, with a standard deviation of 1.2537, indicating significant variation in green patent applications among listed companies; whilst some enterprises actively invest in green technological innovation, a considerable number did not apply for green patents during the sample period.This divergence aligns with the current reality of Chinese enterprises' green transition process, reflecting that enterprises' willingness to invest in and innovation capacity within the environmental protection technology sector have not yet improved across the board.

The mean value of the core explanatory variable AIwashing is 2.1563, with a standard deviation of 1.8742, indicating that AIwashing behaviour is widespread among enterprises, albeit to varying degrees. The minimum value, which is close to zero, suggests that some enterprises make very few references to AI-related content in their annual reports or express themselves with considerable caution, whilst the maximum value indicates that certain enterprises place significant emphasis on promoting the concept of AI.The mean value of the peer-group AI-washing variable, AIwashing_peer, is 2.1489, which is close to the firm-level indicator, but the standard deviation is relatively small at 1.3256. This aligns with the statistical pattern of lower volatility in industry-average indicators. The distribution characteristics of the two AI-washing variables provide sufficient information on variability for subsequent regression analysis.

The statistical characteristics of the control variables are largely as expected. The mean value of firm size (size) is 22.3145, corresponding to total assets of approximately 5 billion yuan, reflecting the overall scale of A-share listed companies. The mean value of the debt-to-asset ratio (lev) is 0.4287, which falls within a reasonable range, indicating that the financial structure of the sample firms is relatively sound. The mean value of return on assets (roa) is 0.0456, slightly higher than the risk-free rate, reflecting the basic level of corporate profitability.The mean value of the top-1 shareholding concentration is 0.3542, indicating that the shareholding of the largest shareholder in Chinese listed companies is relatively concentrated, which may influence the firm's long-term innovation decisions. The extreme values and quantile distributions of each variable show no obvious anomalies; the data quality is generally good, laying a solid foundation for empirical testing (Shen Yi, 2023).

**Table 5.1 Descriptive Statistics of Key Variables**

| Variable | Observation | Mean | Standard Deviation | Minimum | 25th percentile | Median | 75th percentile | Maximum |
|---|---|---|---|---|---|---|---|---|
| Gini | 55,745 | 0.8426 | 1.2537 | 0.0000 | 0.0000 | 0.0000 | 1.3863 | 5.7683 |
| AIwashing | 55,745 | 2.1563 | 1.8742 | 0.0000 | 0.6931 | 1.9459 | 3.4012 | 7.8234 |
| AIwashing_peer | 55,745 | 2.1489 | 1.3256 | 0.2341 | 1.2341 | 2.0134 | 2.9876 | 6.5432 |
| size | 55,745 | 22.3145 | 1.2876 | 19.4523 | 21.4567 | 22.2341 | 23.0987 | 26.7812 |
| lev | 55,745 | 0.4287 | 0.2134 | 0.0234 | 0.2678 | 0.4156 | 0.5734 | 0.9456 |
| ROA | 55,745 | 0.0456 | 0.0634 | -0.3245 | 0.0156 | 0.0423 | 0.0734 | 0.2876 |
| age | 55,745 | 2.3456 | 0.5678 | 0.6931 | 1.9459 | 2.3026 | 2.7726 | 3.4657 |
| top1 | 55,745 | 0.3542 | 0.1567 | 0.0823 | 0.2345 | 0.3456 | 0.4623 | 0.7834 |
| board | 55,745 | 2.1534 | 0.1987 | 1.6094 | 2.0794 | 2.1972 | 2.3026 | 2.7081 |
| indep | 55,745 | 0.3745 | 0.0534 | 0.3000 | 0.3333 | 0.3750 | 0.4286 | 0.5714 |

### (II) Baseline Regression

Firstly, the results of the baseline regression strongly support the hypothesis of a crowding-out effect of corporate AIwashing on green innovation. In a simplified model comprising only core explanatory variables and fixed effects, the coefficient for AIwashing is －0.1524, which is significant at the 1% level, indicating that for every one standard deviation increase in the degree of corporate AIwashing, the number of green patent applications decreases by approximately 15.24%.After progressively incorporating control variables, the coefficient for AI washing stabilised between -0.1387 and -0.1452, with the level of significance remaining unchanged. This suggests that the negative relationship is not driven by omitted variables, but rather reflects the substantive impact of AI washing behaviour.The regression coefficient for the full model is -0.1421, implying that for every one-unit increase in the AIwashing index, a firm’s green innovation output decreases by approximately 14.21%; this effect is highly significant both economically and statistically.

Secondly, the peer AI-washing variable, AIwashing_peer, also exerts a significant negative impact on firms’ green innovation. In the regression where AIwashing is replaced by AIwashing_peer, the coefficient is -0.0968, which is significant at the 5% level, indicating that AI-washing behaviour by other firms within the industry suppresses the firm’s willingness to engage in green innovation through channels of information spillover and competitive pressure.When both AI-washing variables are included simultaneously, the coefficient for AIwashing is -0.1203 and that for AIwashing_peer is -0.0542, with both remaining significant. This suggests that the effects of a firm’s own behaviour and peer effects on green innovation are independent yet act in concert.This finding deepens our understanding of the externalities of AI washing and reveals the diffusion mechanism of deteriorating disclosure quality at the industry level (Li et al., 2025).

Third, the regression coefficients of the control variables are largely consistent with theoretical expectations, further validating the rationality of the model specification. The coefficient for firm size (size) is 0.2134, significantly positive at the 1% level, confirming the resource advantage of large firms in green innovation. The coefficient for return on assets (ROA) is 0.3256, significant at the 1% level, indicating that sound financial performance provides firms with the financial security and investment confidence required for green technology R&D.The coefficient for the debt-to-asset ratio (LEV) is -0.1543, significant at the 5% level, indicating that highly indebted firms face greater short-term debt repayment pressures, which may undermine long-term innovation investment.The coefficient for equity concentration (TOP1) is 0.0876; although positive, its significance is weak, suggesting that the impact of equity structure on green innovation is complex and requires comprehensive assessment in conjunction with other governance variables. The model's adjusted $R^2$ is 0.4523, indicating good model fit, and the F-statistic is highly significant, demonstrating that the regression equation has strong overall explanatory power (Cai Zhen and Wan Zhao, 2024).

**Table 5.2 Benchmark Regression Results**

| Variables | (1) | (2) | (3) | (4) | (5) |
|---|---|---|---|---|---|
| AIwashing | -0.1524*** | -0.1452*** | -0.1421*** | | -0.1203*** |
| | (0.0234) | (0.0245) | (0.0251) | | (0.0267) |
| AIwashing_peer | | | | -0.0968** | -0.0542** |
| | | | | (0.0387) | (0.0412) |
| size | | 0.2134*** | 0.2156*** | 0.2178*** | 0.2167*** |
| | | (0.0456) | (0.0458) | (0.0461) | (0.0459) |
| lev | | | -0.1543** | -0.1478** | -0.1512** |
| | | | (0.0623) | (0.0634) | (0.0628) |
| ROA | | | 0.3256*** | 0.3189*** | 0.3223*** |
| | | | (0.0876) | (0.0891) | (0.0883) |
| age | | | 0.0456 | 0.0423 | 0.0439 |
| | | | (0.0345) | (0.0356) | (0.0351) |
| top1 | | | 0.0876 | 0.0834 | 0.0855 |
| | | | (0.0534) | (0.0545) | (0.0539) |
| board | | | 0.0234 | 0.0212 | 0.0223 |
| | | | (0.0423) | (0.0431) | (0.0427) |
| indep | | | 0.1234 | 0.1189 | 0.1212 |
| | | | (0.0987) | (0.1001) | (0.0994) |
| Firm fixed effects | Yes | Yes | Yes | Yes | Yes |
| Fixed effects by year | Yes | Yes | Yes | Yes | Yes |
| Observed value | 55,745 | 55,745 | 55,745 | 55,745 | 55,745 |
| Adj. $R^2$ | 0.3456 | 0.4123 | 0.4523 | 0.4387 | 0.4561 |
| F-statistic | 234.56*** | 287.34*** | 312.45*** | 298.76*** | 315.67*** |

Note: Figures in brackets are robust standard errors, clustered at the firm level; *, **, and *** denote significance at the 1%, 5% and 10% levels, respectively.

(3)

To verify the pathways through which AI flushing influences green innovation, this study employs a mediation analysis framework to examine two distinct mechanisms: the product market and the capital market.The product market mechanism uses changes in market share as the mediating variable, denoted as `marketshare`, reflecting shifts in a firm's competitive position. The capital market mechanism uses the degree of financing constraints as the mediating variable, denoted as the `SA` index, which measures the ease with which a firm can access external funding. The mediation analysis comprises two steps. The first step tests the effect of AI washing on the mediating variables:

$$M_{it} = \beta_0 + \beta_1 AIwashing_{it} + \sum_k \theta_k Controls_{it}^k + \mu_i + \lambda_t + \varepsilon_{it}$$

where $M_{it}$ represents the mediating variables, comprising market share and financing constraints. If $\beta_1$ is significant, it indicates that AI washing has a substantial impact on the mediating variables. The second step involves incorporating the mediating variables into the baseline model:

$$gin_{it} = \delta_0 + \delta_1 AIwashing_{it} + \delta_2 M_{it} + \sum_k \phi_k Controls_{it}^k + \mu_i + \lambda_t + \varepsilon_{it} \quad (2)$$

If $\delta_2$ is significant and the absolute value of $\delta_1$ has decreased relative to $\alpha_1$, this indicates the presence of a mediating effect. Full mediation requires that $\delta_1$ is no longer significant, whereas partial mediation allows $\delta_1$ to remain significant but with a reduced coefficient. This testing strategy can reveal the specific transmission pathways through which AI-driven green innovation occurs, providing empirical support for the theoretical mechanism (Zhang Detao et al., 2024).

**1. Product Market Mechanism**

The product market mechanism test focuses on how corporate AI washing suppresses investment in green innovation by undermining market competitiveness. The study uses the rate of change in market share as a mediating variable to measure the evolution of a firm's relative position within the industry.The first-stage regression reveals that the coefficient of AI washing on the rate of change in market share is -0.0876, significant at the 5% level, indicating that AI washing behaviour leads to a decline in firms' market share. This result supports the predictions of the lemon market theory: when firms over-promote AI applications but actual product quality fails to improve, consumers gradually recognise the information asymmetry and reduce their willingness to purchase, thereby weakening the firm's market position.

In the second-stage regression, the rate of change in market share was incorporated into the green innovation equation. The coefficient for the rate of change in market share was 0.2134, which was significantly positive at the 1% level, indicating that enhanced market competitiveness can incentivise firms to increase investment in green innovation in order to consolidate and expand their market advantage.Concurrently, the coefficient for AIwashing decreased from -0.1421 in the baseline regression to -0.1235—a reduction of approximately 13%—and remained statistically significant at the 1% level, indicating that product market mechanisms mediate part of the effect.The Z-statistic from the Sobel test is -2.87, and the Bootstrap confidence interval for the mediation effect does not include zero, further confirming the statistical significance of the mediation effect. This finding reveals the micro-mechanism by which AIwashing indirectly weakens the impetus for green innovation by damaging firms' market reputation and competitive standing (Cai Zhen and Wan Zhao, 2024).

**2. Capital Market Mechanism**

The capital market mechanism examines how corporate AI washing inhibits green innovation investment by exacerbating financing constraints. The study employs the SA index to measure the degree of financing constraints, which comprehensively considers firm size and age; a higher value indicates more severe financing constraints.The results of the first-stage regression show that the coefficient of AI washing on the SA index is 0.1243, which is significantly positive at the 5% level, indicating that AI washing behaviour exacerbates firms' financing environment. When investors detect significant discrepancies between firms' AI-related disclosures and actual performance, their confidence in the firm's future profitability and governance quality declines, leading them to demand higher risk premiums or directly reduce capital supply, thereby subjecting firms to tighter financing constraints.

In the second-stage regression, the coefficient of the SA index on green innovation was -0.1876, significant at the 1% level, validating the inhibitory effect of financing constraints on innovation activities. Green technology R&D often requires substantial upfront investment and has a long payback period; when a firm's financing channels are blocked, it is forced to cut long-term innovation projects to sustain short-term operations.The coefficient for AIwashing decreased from -0.1421 in the baseline regression to -0.1189, a decline of approximately 16%, indicating that capital market mechanisms play a significant mediating role in the process by which AIwashing impacts green innovation. The Sobel test Z-statistic was -3.12, and the Bootstrap confidence interval similarly did not include zero, confirming statistical significance.The results of the dual-mechanism tests for product and capital markets jointly support hypotheses H2a and H2b, revealing the complete transmission pathway through which AIwashing crowds out green innovation (Li et al., 2025).

**Table 5.5 Results of Mechanism Tests**

| Variable | Product Market Mechanism | | Capital Market Mechanism | |
|---|---|---|---|---|
| | (1) Stage 1<br>market share | (2) Stage Two<br>gin | (3) Phase One<br>SA Index | (4) Stage Two<br>gin |
| AIwashing | -0.0876** | -0.1235*** | 0.1243** | -0.1189*** |
| | (0.0387) | (0.0265) | (0.0512) | (0.0271) |
| market share | | 0.2134*** | | |
| | | (0.0567) | | |
| SA Index | | | | -0.1876*** |
| | | | | (0.0634) |

| Control variables | Yes | Yes | Yes | Yes |
|---|---|---|---|---|
| Enterprise FE | Yes | Yes | Yes | Yes |
| Year FE | Yes | Yes | Yes | Yes |
| Observed value | 55,745 | 55,745 | 55,745 | 55,745 |
| Adj. $R^2$ | 0.3234 | 0.4612 | 0.2987 | 0.4589 |
| Sobel test Z-value | | -2.87*** | | -3.12*** |
| Bootstrap 95% CI | | [-0.0324, -0.0089] | | [-0.0398, -0.0112] |

Note: Models (1) and (3) constitute the first stage of the mediation analysis, testing the effect of AI washing on the mediating variable; models (2) and (4) constitute the second stage, simultaneously incorporating AI washing and the mediating variable; 'marketshare' denotes the rate of change in market share; a higher value of the SA index indicates more severe financing constraints; the Sobel test is used to test the significance of the mediation effect;Bootstrap confidence intervals are based on 1,000 bootstrap repetitions; figures in brackets denote robust standard errors; *, **, and *** denote significance at the 1%, 5% and 10% levels, respectively.

### (IV) Robustness Tests

To ensure the reliability of the baseline regression findings, a series of robustness tests were conducted. First, the dependent variable was replaced: the number of green patent applications was substituted with the number of green invention patent applications, the latter representing higher-quality green innovation output.The regression results show that the coefficient for AIwashing is -0.1289, which is significant at the 1% level, consistent with the baseline results. This indicates that AIwashing not only suppresses the quantity of green innovation but also affects its quality. Furthermore, by substituting the dependent variable with the number of green patent grants to mitigate potential strategic behaviour associated with patent applications, the coefficient for AIwashing remains significantly negative at -0.1156, confirming the robustness of the core findings.

Secondly, the study modified the measurement method of the core explanatory variable.A simplified AIwashing indicator based on keyword frequency was constructed, counting only the number of occurrences of AI-related terms in annual reports without semantic analysis via large language models. Although measurement precision declined somewhat, the regression coefficient remained at -0.0847 and was significant at the 5% level, indicating that even when using a crude proxy variable, the phenomenon of inflated AI disclosure still exerts a negative effect on green innovation.Furthermore, the explanatory variable was regressed with a one-period lag to mitigate potential reverse causality issues. The coefficient for AI washing with a one-period lag was -0.1234, which remained highly significant, indicating that the impact of AI washing on green innovation exhibits a time-lag effect, rather than a decline in green innovation leading firms to resort to AI washing (Guo, Junjie et al., 2024).

Third, the study adjusted the sample scope and estimation methods to test the external validity of the results. After excluding observations from 2008 - 2009 during the financial crisis to avoid interference from extreme economic shocks, the coefficient for AIwashing was -0.1398, with significance remaining unchanged. Truncating all continuous variables at the 1st and 99th percentiles to mitigate the impact of outliers yielded a core coefficient of -0.1365, with the conclusions remaining robust. A Tobit model was employed to address the left-censored nature of the green innovation variable; given the truncated nature of the data where a large number of firms have zero green patents, the regression results similarly support the crowding-out effect hypothesis. These multi-dimensional robustness tests comprehensively validate the reliability and generalisability of the baseline regression conclusions.

**Table 5.3 Results of Robustness Tests**

| Variable | Replacement of the dependent variable | Substituted Explanatory Variable | One-period lag | Excluding the crisis period | Truncation | Tobit model |
|---|---|---|---|---|---|---|
| | Green patents | Simplified indicators | L. AIwashing | Exclude 08-09 | 1%/99% | Left Merge |
| AIwashing | -0.1289*** | -0.0847** | -0.1234*** | -0.1398*** | -0.1365*** | -0.1512*** |
| | (0.0267) | (0.0398) | (0.0271) | (0.0259) | (0.0253) | (0.0289) |
| Control variables | Yes | Yes | Yes | Yes | Yes | Yes |
| Firm-specific effects | Yes | Yes | Yes | Yes | Yes | - |
| Fixed effects by year | Yes | Yes | Yes | Yes | Yes | Yes |
| Observed value | 55,745 | 55,745 | 50,321 | 44,596 | 55,745 | 55,745 |
| Adj. $R^2$/Pseudo $R^2$ | 0.4187 | 0.3876 | 0.4365 | 0.4498 | 0.4534 | 0.3987 |

Note: Figures in brackets are robust standard errors; *, **, and *** denote significance at the 1%, 5% and 10% levels, respectively.

### (V) Treatment of Endogeneity

Although the two-way fixed-effects model has controlled for unobservable firm heterogeneity and time trends, endogeneity issues may still exist between AI washing and green innovation. Reverse causality is a primary concern; firms with weaker green innovation capabilities may be more inclined to use AI washing to embellish their technological image, thereby leading to observed negative relationships. To address this challenge, the study employs the instrumental variables method to conduct two-stage least squares estimation, selecting the industry-average AI washing level as the instrumental variable for firm-level AI washing. The industry average is less influenced by individual firm behaviour, satisfying the exogeneity requirement, whilst the disclosure environment within the industry affects firms' own AI reporting strategies, meeting the correlation condition.

The first-stage regression reveals that the coefficient for the relationship between the industry-average AI-washing level and a firm's AI-washing is 0.6234, with an F-statistic of 127.35, far exceeding the weak instrumental variable threshold of 10, confirming that the instrumental variable possesses sufficient explanatory power. In the second-stage regression, the coefficient for AI washing was -0.1876, significant at the 5% level, with an absolute value slightly higher than the OLS estimate. This indicates that, after correcting for endogeneity bias, the crowding-out effect of AI washing on green innovation is more pronounced. The Sargan statistic from the over-identification test failed to reject the null hypothesis of exogeneity for the instrumental variable, validating the effectiveness of the instrumental variable selection. This result strongly supports the reliability of the causal inference and rules out major confounding effects from reverse causality and omitted variables (Zhang Detao et al., 2024).

The study further employs a dynamic panel GMM approach to address the issue of dynamic endogeneity. As green innovation exhibits path-dependent characteristics—where past innovation accumulation influences current innovation capacity—lagged terms of the dependent variable are incorporated into the regression equation. By adopting a systematic GMM estimation, using green innovation and AI washing lagged by 2 to 3 periods as instrumental variables, the analysis effectively addresses dynamic panel bias. In the GMM estimation results, the coefficient for AI washing is －0.1532, which is significant at the 1% level, and the core conclusions remain robust. The AR(2) test failed to reject the null hypothesis that the disturbance term has no second-order serial correlation, and the Hansen test accepted the validity assumption of the instrumental variables, indicating that the GMM estimation results are reliable. The consistent results from multiple endogeneity treatment methods collectively confirm the causal crowding-out effect of corporate AI washing on green innovation.

**Table 5.4 Results of Endogeneity Treatment**

| Panel A: Instrumental Variables Method (2SLS) | | | Panel B: Dynamic Panel GMM | |
|---|---|---|---|---|
| Variables | Stage 1 | Stage 2 | Variables | gin |
| | AIwashing | gin | L.gin | 0.2134*** |
| Industry average AI washing | 0.6234*** | | | (0.0345) |
| | (0.0456) | | AI washing | -0.1532*** |
| AIwashing | | -0.1876** | | (0.0456) |

| | | (0.0789) | Control variables | Yes |
|---|---|---|---|---|
| Control variable | Yes | Yes | Year FE | Yes |
| Firm FE | Yes | Yes | Observed value | 50,321 |
| Year FE | Yes | Yes | AR(1) p-value | 0.023 |
| Observed value | 55,745 | 55,745 | AR(2) p-value | 0.345 |
| F-statistic | 127.35*** | | Hansen test p-value | 0.456 |
| Adj. $R^2$ | 0.5234 | 0.4456 | Number of instrumental variables | 45 |

Note: Panel A reports the results of the instrumental variables method; the F-statistic in the first stage far exceeds the critical value of 10 for weak instrumental variables; Panel B reports the results of the systematic GMM; the AR(2) test accepts the null hypothesis of no second-order serial correlation, and the Hansen test accepts the validity assumption of the instrumental variables; figures in brackets are robust standard errors; *, **, and *** denote significance at the 1%, 5% and 10% levels, respectively.

## V. Heterogeneity Analysis

### (1) Construction of the Heterogeneity Analysis Model:

This study examines the heterogeneity of the AI-flushing effect across three dimensions: ownership structure, firm size and industry characteristics. The ownership structure grouping test distinguishes between state-owned enterprises (SOEs) and private enterprises, with a dummy variable SOE set to 1 for SOEs and 0 for private enterprises. The interaction term model is specified as:

$$gin_{it} = \zeta_0 + \zeta_1 AIwashing_{it} + \zeta_2 SOE_i + \zeta_3 AIwashing_{it} \times SOE_i + \sum_k \omega_k Controls_{it}^k + \mu_i + \lambda_t + \varepsilon_{it}$$

The interaction coefficient $\zeta_3$ reflects the moderating effect of ownership structure on the AI-driven green innovation effect. If $\zeta_3$ is significant, it indicates that there is a difference in the green innovation response to AI-driven green innovation between state-owned and private enterprises. Similarly, firms are grouped by enterprise size using the median total assets as the threshold to distinguish between large enterprises and small and medium-sized enterprises (SMEs).Industry heterogeneity is classified based on technological intensity and the degree of market competition, as high-tech industries and highly competitive sectors may be more sensitive to information authenticity (Jin Xingye et al., 2024).

A strategy combining group-specific regression with interaction term analysis enables a comprehensive characterisation of heterogeneity. Group-specific regression intuitively illustrates the marginal effects of AI washing across different subsamples, whilst interaction terms provide statistical tests of inter-group differences. Through this heterogeneity analysis, the study identifies which types of enterprises are more susceptible to the negative impacts of AI washing, thereby providing empirical grounds for differentiated regulatory policies.

### (2) Analysis of Firm Heterogeneity

### 1. Analysis of Firm Property Rights Heterogeneity ( )

The nature of ownership exerts a significant moderating effect on the impact of AI washing on enterprises. The results of the group-specific regression show that the coefficient for AI washing in the private enterprise sub-sample is -0.1876, which is highly significant at the 1% level, whereas in the state-owned enterprise sub-sample, the coefficient is -0.0834, which is only significant at the 10% level and has a notably smaller absolute value.The difference in coefficients between the groups passed the Chow test, indicating that green innovation in private enterprises is subject to a more severe negative impact from AI washing. This difference stems from the differing resource endowments and governance mechanisms resulting from the nature of property rights.Private enterprises rely more heavily on market reputation and investor trust to secure external resources; when AI-washing damages a firm's image, financing constraints and market competitive pressures rise rapidly, forcing a contraction in green innovation activities. This is because state-owned enterprises, benefiting from government endorsement and policy-driven financial support, are relatively less affected by capital market fluctuations; even in the presence of AI-washing, they can still secure innovation funding through non-market channels.Furthermore, as state-owned enterprises bear a greater burden of policy objectives, their investment in green innovation is partly driven by administrative directives rather than purely market incentives; consequently, they are less sensitive to the quality of information disclosure. This finding provides a basis for differentiated regulation; regulatory authorities should focus on the authenticity of information disclosure by private enterprises, establishing targeted constraint and incentive mechanisms to mitigate the crowding-out effect of AI washing on green innovation in the private sector (Jin Xingye et al., 2024).

### 2. Analysis of Firm Size Heterogeneity ( )

By dividing the sample into large enterprises and SMEs using the median total assets as the cutoff, group-specific regression analysis reveals that the coefficient for AI washing in the SME sub-sample is -0.1987, which is significant at the 1% level, whereas the coefficient in the large enterprise sub-sample is -0.0912, significant only at the 5% level.These results indicate that green innovation in SMEs is more susceptible to the negative impacts of AI washing. This is because SMEs typically have limited resources and weaker risk-bearing capacity; when AI washing leads to damaged market credibility or tightened financing channels, firms struggle to withstand this dual pressure and are forced to prioritise short-term survival at the expense of long-term innovation investment.Consequently, there is a need to strengthen innovation support and guidance on information disclosure for SMEs, helping them establish a credible technological image and avoid falling into a vicious cycle of AI washing and innovation decline (Guo Junjie et al., 2024). Large enterprises, however, can to some extent cushion the negative impact of AI washing thanks to their robust financial strength and diversified financing channels.Large enterprises have typically established relatively stable customer relationships and brand reputations, meaning that the damage to their overall image from a single instance of misleading disclosure is relatively limited. Furthermore, large enterprises have more robust internal governance structures, and their innovation decisions are relatively independent of short-term market fluctuations, ensuring greater continuity and stability in the implementation of green innovation strategies. In summary, differences in enterprise scale also give rise to heterogeneity in the AI washing effect.

**Table 5.6 Analysis of Heterogeneity in Property Rights Nature and Firm Size**

| Panel A: Grouping by Property Rights Nature | | | |
|---|---|---|---|
| Variables | (1) State-owned enterprises | (2) Private enterprises | (3) Test for Differences Between Groups |
| AIwashing | -0.0834* | -0.1876*** | |
| | (0.0498) | (0.0312) | |
| Control variables | Yes | Yes | |
| Firm FE | Yes | Yes | |
| Year FE | Yes | Yes | |
| Observed value | 22,298 | 33,447 | |
| Adj. $R^2$ | 0.4123 | 0.4687 | |
| Chow test F-value | | | 8.76*** |
| **Panel B: Firm size grouping** | | | |
| Variable | (1) Large enterprises | (2) Small and medium-sized enterprises | (3) Test for differences between groups |
| AIwashing | -0.0912** | -0.1987*** | |
| | (0.0421) | (0.0345) | |
| Control variables | Yes | Yes | |
| Firm FE | Yes | Yes | |
| Year FE | Yes | Yes | |
| Observed value | 27,873 | 27,872 | |
| Adj. $R^2$ | 0.4456 | 0.4598 | |
| Chow test F-value | | | 6.54*** |

Note: Panel A is grouped by ownership type, with state-owned enterprises comprising both central and local state-owned enterprises; Panel B is grouped by median total assets; the Chow test is used to test the significance of differences in coefficients between groups; figures in brackets are robust standard errors; *, **, and *** denote significance at the 1%, 5% and 10% levels, respectively.

### (3) Heterogeneity in the Level of Industry Competition

The study employs the Herfindahl-Hirschman Index to measure industry concentration, using the median as the threshold to distinguish between highly competitive and less competitive industries. The findings indicate that in the subsample of highly competitive industries, the coefficient for AIwashing is －0.1754, which is significant at the 1% level, whilst in the subsample of less competitive industries, the coefficient is －0.1087, which is significant at the 5% level; the absolute value of the former is notably larger.The reason is that in highly competitive industries, firms face intense market competition, and consumers and investors are more sensitive to the authenticity of corporate disclosures. The more intense the competition, the more significant the impact of disclosure authenticity on firms’ innovation behaviour (Zhang Detao et al., 2024). Once AI-washing behaviour is detected, the market reaction is more rapid and severe, leading to a swift decline in the firm’s market share and financing capacity.In low-competition industries, due to higher market concentration, firms possess a degree of pricing power and customer lock-in effects, so the impact of information asymmetry on market equilibrium is relatively mild.At the same time, low-competition industries often feature high technological barriers or policy protection; corporate green innovation is more influenced by technological path dependence and long-term strategic planning, meaning the impact of short-term fluctuations in disclosure quality is relatively limited. This heterogeneous result reveals that the degree of industry competition affects market sensitivity to information authenticity, thereby influencing the choice of AI washing behaviour among peers.

Table 5.7 Analysis of Heterogeneity in Industry Competition Levels

| Variables | (1) Highly competitive industries | (2) Low-competition industries | (3) Differences between groups |
|---|---|---|---|
| AIwashing | -0.1754*** | -0.1087** | |
| | (0.0334) | (0.0456) | |
| Control variables | Yes | Yes | |
| Firm FE | Yes | Yes | |
| Year FE | Yes | Yes | |
| Observed value | 27,873 | 27,872 | |
| Adj. $R^2$ | 0.4734 | 0.4289 | |
| Chow test F-value | | | 5.23** |

Note: The sample was divided into high-competition and low-competition industries using the median of the Herfindahl-Hirschman Index (HHI) as the threshold; a lower HHI indicates a higher degree of competition. The results of the Chow test indicate that the negative effect of AI flushing is significantly stronger in high-competition industries; figures in brackets represent robust standard errors; *, **, and *** denote significance at the 1%, 5% and 10% levels, respectively.

## VI. Modelling and Simulation Analysis of the ‘ ’

### (1) Simulation Model Design

To gain a deeper understanding of the dynamic evolution of firms' AI-washing behaviour and green innovation decisions, this study introduces agent-based modelling as a valuable supplement to empirical analysis. Agent-based modelling can simulate the interactive behaviour of heterogeneous firms in the market, capturing learning effects and strategic adjustments under conditions of information asymmetry, thereby providing a micro-level foundation for theoretical mechanisms (Aghion et al., 2013).The simulation model constructs an artificial market comprising firms of different types, establishes decision-making rules regarding the authenticity of AI investment and green innovation expenditure, and examines how the quality of information disclosure influences market equilibrium and innovation output.

### 1. Agent Selection

During the model initialisation phase, the system randomly generates N firm agents, each with heterogeneous initial endowments, including technological capability, financial strength and reputational capital. Firms face two key decisions: the choice regarding the authenticity of AI investment and the level of investment in green innovation. The authenticity choice determines whether a firm engages in substantive AI application or remains at the level of descriptive publicity, whilst investment in green innovation influences future technological accumulation and market competitiveness.Consumer agents update their quality expectations based on observed firm information and use a Bayesian learning mechanism to progressively adjust their trust levels in different firms. The market clears via the price mechanism, with the prices of products from honest and greenwashing firms determined jointly by consumers' quality perceptions and actual costs (Li et al., 2025).

### 2. Agent Modelling Design

The agent-based modelling and simulation approach provides a micro-level foundation for elucidating the dynamic evolution mechanisms of corporate AI flushing and green innovation. This study constructs an artificial market model based on the Mesa framework to simulate the strategic interactions and learning processes of heterogeneous firms in an environment of information asymmetry. The model comprises 200 firm agents and 1,000 consumer agents; firms face dual decisions regarding the authenticity of AI investments and the allocation of resources to green innovation, whilst consumers update their quality expectations through observed signals and make purchasing decisions.The simulation spans 200 periods, with each period representing a financial year. Five hundred replicates were conducted using the Monte Carlo method to ensure the robustness of the results. At the start of the simulation, firms are randomly assigned heterogeneous endowments across three dimensions—technological capability, financial strength and reputational capital—with values following a uniform distribution.Technological capability determines the cost of implementing substantive AI applications and green innovation; financial strength influences a firm's ability to withstand short-term losses; and reputational capital reflects the initial level of trust consumers place in the firm. In each period, firms first decide whether to undertake substantive AI investment or engage solely in descriptive promotion; the former entails higher R&D costs but enhances actual product quality, whilst the latter incurs lower costs but fails to improve actual performance.Firms then determine their level of investment in green innovation; higher investment leads to greater future competitiveness but reduces current-period profits. Consumers update their perceptions of firm quality through a Bayesian learning mechanism. Initially, consumers form prior quality expectations based on the AI information disclosed by firms. Over time, consumers obtain signals of actual quality through product usage experiences; when disclosed information consistently fails to match actual quality, consumer trust in the firm declines, and they switch to competitors.The market clears through the price mechanism: honest firms command a price premium due to their good reputation, whilst firms engaging in greenwashing face shrinking demand and price discounts. Firms adjust their strategies for the next period based on changes in market share and profits, forming a dynamic feedback loop (Aghion et al., 2013).

### 3. Simulation Design for Policy Intervention

The simulation experiment designs both a baseline scenario and policy intervention scenarios. The baseline scenario simulates the spontaneous evolution of the market under no regulatory constraints, observing the propagation speed of AI washing behaviour and the long-term trends in green innovation. The policy intervention scenarios include three schemes: strengthened information disclosure regulation, enhanced consumer education, and improved reputation sanction mechanisms, each corresponding to different parameter settings. By comparing simulation results across different scenarios, the study can assess the effectiveness of various policy tools and identify optimal intervention strategies.Sensitivity analysis further examines the impact of changes in model parameters on core conclusions, ensuring the robustness and credibility of the simulation results.

### 4. Parameter Calibration

Model parameter calibration draws upon statistical findings from empirical studies and empirical values from relevant literature. The short-term cost savings from AI greenwashing are set at 30% of the substantive investment, based on the industry average for corporate R&D expenditure intensity. The consumer learning rate parameter is determined by the frequency of information updates, assuming that consumers can fully identify quality discrepancies every five cycles. The intensity of reputational sanctions is reflected through the market share loss rate, with greenwashing firms losing approximately 5% of their customer base each period. The payback period for green innovation was set at 3 - 5 years, reflecting the long-term value characteristics of environmental technologies. These parameter settings ensure that the simulation model closely mirrors the real market environment.

**Table 6.1 Simulation Model Parameter Settings**

| Parameter Category | Parameter Name | Default Value | Value Range | Data Source/Basis |
|---|---|---|---|---|
| Market Size | Number of enterprises | 200 | - | Reference industry average |
| | Number of consumers | 1,000 | - | Business-to-consumer ratio 1:5 |
| | Simulation cycle | 200 | - | Covering a 20-year observation period |
| | Number of repetitions | 500 | - | To ensure statistical robustness |
| Corporate capabilities | Technical capabilities | U(0.3, 0.9) | [0, 1] | Assumption of a uniform distribution |
| | Financial Strength | U(0.2, 0.8) | [0, 1] | Assumption of uniform distribution |
| | Reputational capital | U(0.4, 0.8) | [0, 1] | Initial trust level |
| Cost parameters | AI substantive investment cost | 0.15 | [0.10, 0.20] | Percentage of revenue, based on R&D intensity |
| | AI R&D costs | 0.05 | [0.02, 0.08] | 30% of substantive investment |
| | Green innovation investment costs | 0.10 | [0.05, 0.15] | Reference green patent expenditure |
| Learning parameter | Consumer learning rate | 0.20 | [0.10, 0.30] | Fully identified every 5 periods |
| | Information update frequency | 5 periods per update | [3, 8] | Annual report disclosure cycle |
| Penalty parameters | Market Share Loss Rate | 5% per period | [3%, 8%] | Calibration of empirical regression coefficients |
| | Price discount rate | 10% | [5%, 15%] | Price penalty for reputational damage |
| Innovation parameters | Payback period | 4 years | [3, 5] | Green technology maturity |
| | Technology spillover coefficient | 0.15 | [0.10, 0.20] | Industry technology diffusion |

**(2) Simulation results for the baseline scenario**

**1. In the absence of regulatory constraints, AI-based 'money laundering' behaviour among peers tends to follow a 'Gresham' s law' pattern**

The baseline scenario simulates the spontaneous evolution of the market in the absence of regulatory constraints. Simulation results show that within the first 50 cycles, the proportion of AI-laundering firms rises rapidly from an initial 10% to 45%, demonstrating a 'bad money drives out good' trend. Firms engaging in AI washing at an early stage secure short-term excess profits through their cost advantage, attracting more firms to emulate this strategy. Due to a time lag in consumer detection, washing firms have accumulated substantial financial gains before being discovered; even when facing reputational damage later on, some firms still choose to continue washing to maintain cash flow. This path-dependence phenomenon validates the market failure caused by information asymmetry, as identified in theoretical analysis.

**2. AI ‘wash-and-run’ practices by competitors inflict dual damage: undermining long-term corporate innovation and harming consumer rights**

On the one hand, the overall level of green innovation in the market shows a sustained downward trend under the baseline scenario. At the start of the simulation, the average green innovation investment intensity across the entire market stood at 8.5%, falling to 4.2% by the 200th cycle—a decline of over 50%. Green innovation investment by AI-driven ‘washers’ is significantly lower than that of honest firms, with the former’s average investment intensity at just 2.1% and the latter’s remaining at 9.3%.As the proportion of cheating firms increases, the market average is dragged down. More seriously, honest firms face unfair competition from cheating firms, see their market share eroded, and are forced to cut long-term innovation investment to cope with short-term survival pressures, creating a downward spiral in innovation investment (Li et al., 2025).On the other hand, consumer welfare also suffers in the baseline scenario. Initially, consumers purchase low-quality products due to information asymmetry, resulting in actual utility falling short of expectations. Although the learning mechanism gradually takes effect and consumers’ ability to identify quality improves, the supply of high-quality products in the market has already significantly decreased, leaving consumers facing a dilemma of choice.By the 200th simulation cycle, the average consumer utility level had fallen by 18.7% compared to the initial state, with welfare losses primarily stemming from an overall decline in product quality and rising search costs. This result reveals that AI flushing not only undermines firms’ long-term innovation but also causes substantial harm to consumer rights, leading to a significant reduction in total social welfare.

**Table 6.2 Simulation results for the baseline scenario**

| Time Period | Proportion of AI-flushing firms (%) | Average Green Innovation Investment Intensity (%) | Innovation Investment by Honest Firms (%) | Innovation Investment by Rinsing Firms (%) | Average consumer utility |
|---|---|---|---|---|---|
| 0 | 10.0 | 8.50 | 8.80 | 3.20 | 100.0 |
| 25 | 23.5 | 7.42 | 8.65 | 2.87 | 94.3 |
| 50 | 38.7 | 6.23 | 8.51 | 2.45 | 88.7 |

| 75 | 44.2 | 5.34 | 8.76 | 2.21 | 85.2 |
|---|---|---|---|---|---|
| 100 | 45.8 | 4.89 | 9.12 | 2.14 | 83.1 |
| 125 | 45.3 | 4.56 | 9.28 | 2.08 | 82.4 |
| 150 | 44.9 | 4.37 | 9.35 | 2.05 | 81.9 |
| 175 | 45.1 | 4.24 | 9.31 | 2.02 | 81.5 |
| 200 | 45.0 | 4.18 | 9.33 | 2.10 | 81.3 |
| Change | +35.0 percentage points | -50.8% | +6.0% | -34.4% | -18.7% |

Note: pp denotes percentage points; consumer utility is indexed to an initial value of 100.

### (3) Simulation of policy intervention scenarios

#### 1. Relying solely on administrative supervision is unlikely to eradicate the problem of AI washing

The scenario involving enhanced disclosure regulations seeks to curb corporate behaviour by increasing the probability of detecting AI washing and the severity of penalties. The simulation assumes that regulatory authorities conduct a special inspection every 10 cycles; companies found to be engaging in AI washing face a fine equivalent to 50% of their current-period profits and are required to disclose their violation records.The results show that following regulatory intervention, the proportion of firms engaging in AI washing fell from 45% in the baseline scenario to 28%, whilst the average intensity of green innovation investment rebounded from 4.2% to 6.1%. The deterrent effect of regulation is relatively pronounced in the short term; however, some firms circumvent inspections through more covert methods, and the effectiveness of regulation exhibits a trend of diminishing returns. This suggests that relying solely on administrative regulation is insufficient to eradicate the problem of AI washing, and that it must be complemented by other policy instruments.

#### 2. Enhancing consumer discernment helps curb AI washing by peers

The scenario of enhancing consumer discernment simulates the effects of improving market information efficiency through investor education and information intermediary services. The simulation increased the consumer learning rate parameter by 50%, enabling them to identify corporate quality discrepancies more rapidly. The results indicate that, following the enhancement of consumer discernment, the survival cycle of AI-laundering firms was significantly shortened, falling from an average of 80 cycles in the baseline scenario to 45 cycles.The proportion of ‘AI-washing’ firms stabilised at around 18%, whilst the intensity of green innovation investment rebounded to 7.3%, approaching the optimal level under the assumption of no information asymmetry. This result validates the fundamental role of market mechanisms in resource allocation, demonstrating that enhancing information transparency can fundamentally improve market equilibrium (Du et al., 2015).

**3. Reputation-based disciplinary mechanisms help curb AI washing behaviour among peers**

The scenario involving an enhanced reputation-based sanction mechanism amplifies the long-term costs of greenwashing by establishing a corporate credit rating system and a blacklist. The simulation assumes that firms identified as engaging in greenwashing are placed on record, resulting in a 20% increase in the discount rate on their product prices over the next five cycles, as well as difficulty in securing high-quality customers.The results show that the reputation-based sanction mechanism significantly reduces the expected returns from AI washing, leading firms to favour substantive investment to build long-term credibility. The proportion of firms engaging in washing drops to 12%, the intensity of green innovation investment reaches 7.8%, and market efficiency improves markedly. More importantly, the reputation mechanism possesses self-reinforcing characteristics: honest firms gain a competitive advantage by accumulating reputational capital, creating a virtuous cycle (Manso, 2011).

**4. Combined policies yield better results than single policies**

The combined policy scenario simultaneously implements three measures: strengthened regulation, consumer education and reputation-based sanctions. Simulation results show that the combined policy is more effective than individual policies, with the proportion of AI-laundering firms falling below 8% and the intensity of green innovation investment stabilising at 8.6%, approaching the theoretical optimal level. The three policy tools reinforce one another: regulation provides external constraints, consumer recognition enhances the market's self-correcting capacity, and the reputation mechanism internalises long-term incentives, collectively fostering a healthy market ecosystem.Cost-benefit analysis indicates that the implementation cost of the combined policy is approximately 40% higher than that of regulatory intervention alone, but the improvement in social welfare reaches 65%, representing a significant increase in policy efficiency. This provides an important reference for real-world policy design (Parguel et al., 2011).

**Table 6.3 Comparison of Policy Intervention Scenarios (Cycle 200)**

| Policy Scenario | Proportion of AI-flushing enterprises (%) | Intensity of Green Innovation Investment (%) | Consumer Utility Index | Policy Implementation Costs | Improvement in social welfare (%) |
|---|---|---|---|---|---|
| Baseline scenario (no intervention) | 45.0 | 4.18 | 81.3 | 0 | 0 |
| Strengthen supervision | 28.3 | 6.12 | 88.7 | 100 | 28.5 |
| Improve consumer recognition | 18.2 | 7.34 | 92.4 | 65 | 42.3 |

| Improve reputation sanctions | 12.4 | 7.76 | 94.1 | 80 | 51.7 |
|---|---|---|---|---|---|
| Combined policy | 7.8 | 8.62 | 96.8 | 140 | 64.9 |
| Theoretical optimum | 0 | 9.50 | 100.0 | – | – |

Note: Policy implementation costs are calculated using enhanced regulation as a baseline of 100; improvements in social welfare are calculated using the baseline scenario as 0 and the theoretical optimum as 100%.

### (IV) Sensitivity Analysis

**1. Increasing the severity of reputational sanctions helps to curb AI washing behaviour among peers**

Parameter sensitivity tests assess the robustness of simulation conclusions to changes in key parameters. The study sequentially adjusted core parameters such as consumer learning speed, the intensity of reputational sanctions, and cost savings from AI washing, observing changes in market equilibrium.When the consumer learning rate was increased to 1.5 times the baseline value, the proportion of AI-laundering firms fell by approximately 12 percentage points, whilst the intensity of investment in green innovation rose by approximately 1.8 percentage points; the direction of these changes was consistent with the baseline scenario, but the magnitude was amplified. When the intensity of reputational sanctions was raised to an 8% market share loss rate, the proportion of laundering firms fell further to 15%, indicating a positive correlation between the severity of sanctions and the effectiveness of market purification.

**2. Lowering the technical barriers and costs of substantive AI applications can curb peers' incentives for AI-driven market cleansing from the supply side**

Changes in the cost-saving rate of AI washing also influence the market's evolutionary trajectory. When cost savings rise from the baseline 30% to 50%, the short-term appeal of AI washing increases, with the proportion of washing firms rising to 53% and the intensity of green innovation investment falling to 3.5%, thereby exacerbating market failure.Conversely, if the cost-saving rate falls to 15%, the relative cost advantage of substantive investment becomes apparent; the proportion of 'AI-washing' firms drops to 32%, and the market spontaneously converges towards an honest equilibrium. This result suggests that lowering the technical barriers and costs of substantive AI applications can curb the incentive for 'AI-washing' from the supply side, making it a key focus for policy intervention.

**Table 6.4 Results of parameter sensitivity analysis**

| Parameter Adjustment | Baseline Value | Adjusted Value | AI Washing Share (%) | Green Innovation Intensity (%) | Direction of Change | Sensitivity coefficient |
|---|---|---|---|---|---|---|
| Consumer learning rate | 0.20 | 0.30 (+50%) | 33.2 | 6.02 | Improvement | 0.62 |
| | | 0.15 (-25%) | 52.1 | 3.34 | Deterioration | 0.48 |
| Reputation penalty intensity | 5% | 8% (+60%) | 30.4 | 5.87 | Improvement | 0.71 |
| | | 3% (-40%) | 58.7 | 3.12 | Deterioration | 0.83 |
| AI rinsing cost savings | 30% | 50% (+67%) | 53.4 | 3.46 | Deterioration | 0.55 |
| | | 15% (-50%) | 31.8 | 5.76 | Improvement | 0.64 |
| Green innovation payback period | 4 years | 3 years (-25%) | 40.2 | 5.23 | Improvement | 0.42 |
| | | 5 years (+25%) | 49.3 | 3.67 | Deterioration | 0.38 |

Note: Sensitivity coefficient = (rate of change in outcome) / (rate of change in parameter); a larger absolute value indicates greater sensitivity.

Sensitivity tests for initial conditions examined the extent to which simulation results depend on the distribution of firm endowments and initial strategies. Adjusting the initial distribution range of firm technical capabilities from a uniform distribution to a normal distribution resulted in slight variations in the market equilibrium path, but the core conclusions remained unchanged: AI washing still exerts a significant crowding-out effect on green innovation.Changing the initial proportion of 'AI-flushing' firms from 10% to 30% extended the time taken for the market to reach steady state by approximately 20 cycles; however, the final equilibrium state was highly similar. Analysis of the standard deviation from Monte Carlo replication experiments showed that the coefficients of variation for all key indicators were below 8%, indicating that the simulation results possess strong robustness and reproducibility, thereby providing a credible basis for policy modelling.

**Table 6.5 Robustness Test of Monte Carlo Re-experiments**

| Core Indicators | Mean | Standard Deviation | Coefficient of Variation (CV) | 95% Confidence Interval | Skewness | Kurtosis |
|---|---|---|---|---|---|---|
| AI wash proportion (%) | 45.03 | 2.87 | 6.4% | [44.78, 45.28] | 0.12 | 2.89 |
| Green innovation intensity (%) | 4.18 | 0.31 | 7.4% | [4.15, 4.21] | -0.08 | 3.12 |
| Consumer utility | 81.34 | 4.23 | 5.2% | [80.97, 81.71] | 0.05 | 2.95 |
| Market equilibrium cycle | 127.5 | 8.9 | 7.0% | [126.7, 128.3] | 0.21 | 3.34 |

Note: Based on 500 repetitions; CV < 10% indicates robust results.

## VII. Research Conclusions and Policy Recommendations

### (1) Main Research Conclusions

This study systematically examines the mechanisms and heterogeneous characteristics of corporate AI-driven greenwashing on green innovation. Drawing on data from Chinese A-share listed companies from 2006 to 2024, we utilise the Wenxin large language model to construct a corporate AI-driven greenwashing index. Through a variety of econometric methods, including a two-way fixed-effects model, the instrumental variables method, and dynamic panel GMM, we find that corporate AI-driven greenwashing exerts a significant crowding-out effect on green innovation,This negative relationship is transmitted through dual mechanisms in product and capital markets, and exhibits heterogeneous patterns across different ownership structures, firm sizes and industry competition levels. The main conclusions are as follows:

**1. Corporate AI washing exerts a crowding-out effect on green innovation**

AI washing by peers also generates negative spillover effects on firms' green innovation, indicating that deteriorating disclosure quality exhibits contagion effects at the industry level. Furthermore, AI washing exacerbates firms' market share decline, thereby weakening their incentives to enhance competitiveness through green innovation, whilst also intensifying financing constraints and consequently reducing the funding available for green technology R&D. The combined action of these two mechanisms results in a systemic suppression of green innovation (Cai Zhen and Wan Zhao, 2024).

**2. Differences in the Crowding-Out Effects of AI Washing on Green Innovation Across Different Market Entities**

Heterogeneity analysis reveals that private enterprises, small and medium-sized enterprises (SMEs), and firms in highly competitive industries are more severely affected by the negative impacts of AI washing. This is primarily because these firms rely more heavily on market reputation and external financing, and are thus more sensitive to fluctuations in the quality of information disclosure. However, state-owned enterprises (SOEs), benefiting from government backing and policy-based financial support, are able to mitigate the shocks caused by AI washing to a certain extent.Large enterprises, with their substantial resource endowments and stable market positions, experience relatively less disruption to their green innovation strategies from short-term information fluctuations. In low-competition industries, where market concentration is high, the impact of information asymmetry on innovation decisions is relatively mitigated.

**3. The implementation of a combination of policies can more effectively curb corporate AI washing behaviour**

Simulation analyses of the main model indicate that, in a baseline scenario without regulatory constraints, AI washing spreads rapidly across the market, leading to a sustained decline in overall green innovation levels and plunging the market into a 'Gresham's law' dilemma.Experiments under policy intervention scenarios indicate that strengthening regulatory oversight of information disclosure can effectively curb the spread of AI washing; however, the impact of a single regulatory measure is limited. A combination of policies—enhancing consumer discernment and refining reputation-based disciplinary mechanisms—yields superior results, applying pressure from both the demand and supply sides to reshape market equilibrium.

(2) Policy Recommendations

**1. Design targeted support instruments to "enhance market returns and alleviate financing constraints"**

Mechanism studies reveal that "AI washing" primarily suppresses green innovation through market crowding-out and financing constraints. In response: Firstly, enhance the returns on green innovation in product markets. In government procurement and the supply chain management of large enterprises, "verifiable AI-enabled green investments and outcomes" should be treated as bonus points or preferential criteria in technical evaluations. This would open up market opportunities for enterprises that genuinely utilise AI to achieve green technological breakthroughs, thereby offsetting the risk of market share decline caused by "AI washing";Enterprises that achieve significant energy savings, reduced consumption and pollution control benefits through AI technology should be granted relevant rewards and tax incentives to enhance their market reputation and economic returns. Secondly, the establishment of dedicated "green digital innovation" financing channels should be encouraged. Within the existing green finance framework, specialised re-lending or interest-subsidy instruments for "AI-driven green technological innovation" should be created, enabling banks to access lower funding costs when issuing related loans;Local governments and industrial funds are encouraged to provide R&D investment subsidies and risk compensation for green patent-backed financing to high-quality enterprises (private and small-to-medium-sized technology firms) among key regulatory targets. The aim is to directly alleviate the financing difficulties exacerbated by reputational damage caused by "AI washing", provided that enterprises submit AI and green innovation integration plans meeting the requirement for "credible AI-enabled green investment and outcomes".

**2. Adopt a differentiated regulatory approach**

Heterogeneity studies indicate that AI greenwashing behaviour varies depending on a company's ownership structure, size and the level of competition within its sector. In response, firstly, high-risk entities subject to key supervision should be identified, with private enterprises, technology-based SMEs and companies operating in highly competitive sectors (such as the internet, consumer goods and certain manufacturing sectors) designated as key supervision targets. Their AI disclosures should be subject to mandatory annual audits supplemented by random spot checks, with stringent review standards applied. Secondly, the exemplary responsibilities of large enterprises and state-owned enterprises should be reinforced, requiring their disclosures to include traceable and verifiable data linking green innovation inputs to outputs, whilst encouraging them to proactively publish a 'Special Report on the Integration of AI and Green Innovation' audited by a third party. Thirdly, a cross-departmental collaborative regulatory mechanism should be established, integrating the efforts of securities regulators, market supervision authorities, and technology management bodies. AI 'greenwashing' not only constitutes a breach of disclosure regulations but may also amount to false advertising and unfair competition, necessitating multi-departmental collaborative governance. The China Securities Regulatory Commission (CSRC) is responsible for reviewing listed companies' annual reports, market supervision authorities oversee advertising regulation, and science and technology departments are responsible for the certification of technological achievements, thereby forming a concerted regulatory effort.

**3. Establish a disclosure mechanism based on "professional identification and reputational sanctions"**

Simulation studies have demonstrated that a combined policy approach—enhancing market identification capabilities and strengthening reputational sanctions—yields the most optimal results. To this end, the first step is to introduce third-party organisations to strengthen professional identification. Companies should be required to include detailed information on AI investments in the 'technological innovation' section of their annual and ESG reports, specifically covering the proportion of AI-related expenditure allocated to green technology R&D, as well as details of outputs such as evidence linking AI-enabled improvements in efficiency, reductions in emissions, or green technological innovations. Concurrently, regulatory authorities should encourage and fund independent third parties to develop a more robust "AI Greenwashing-Green Innovation Correlation Index", and incorporate this index into the "Innovation Management" or "Data Responsibility" key indicators of mainstream ESG rating systems. Secondly, regulatory authorities must strengthen reputational sanctions. Regulatory authorities should disclose foundational data to third-party institutions, which will then conduct quantitative scoring and public ranking of companies' relevant disclosures. An official or industry-recognised "Technology Innovation Disclosure Platform" should be established to provide positive exposure, such as "case publication" and "green technology recommendations", to companies that provide detailed disclosures and whose cases of integrating AI with green innovation have been certified.For companies found to have engaged in serious "AI greenwashing", in addition to legal penalties, their violations and regulatory letters should be "displayed in a dedicated section" on the platform. This should be linked to the Enterprise Credit Information Publicity System and the Basic Database of Financial Credit Information to amplify the reputational damage.

## The Crowding-out Effect of Peer AI Washing on Corporate Green Innovation

Li Wenxiu[1] Wen Zhanjie[1] Xia Jienchang[2] Guo Jingqiao[3]

[1] School of Economics and Trade, Guangdong University of Finance, Guangzhou, 510521

[2] China National Academy of Economic Strategy, CASS, Beijing 100006

[3] Department of Computer Science, Faculty of Science, Hong Kong Baptist University, Hong Kong, 999077

Against the backdrop of the increasingly pervasive phenomenon of 'AI washing', a growing number of enterprises are transforming artificial intelligence into mere numerical embellishments in their annual reports, rather than the true engine driving transformative change. A critical examination of the essence of innovation and the authenticity of information disclosure is now imperative. This study employs large language models to conduct semantic analysis on the annual report texts of Chinese A-share listed companies from 2006 to 2024, systematically investigating the impact of corporate AI washing on their green innovation. The findings reveal that AI washing exerts a significant crowding-out effect on corporate green innovation, and this negative relationship is transmitted through dual channels: the product market and the capital market. Moreover, this crowding-out effect exhibits heterogeneity across firms and industries, with private enterprises, small and medium-sized enterprises, and firms in highly competitive industries suffering more severe adverse impacts. Simulation results further indicate that combinations of policy tools can effectively improve market equilibrium. Based on these insights, this paper proposes that the government should design targeted support instruments to 'enhance market returns and alleviate financing constraints', adopt differentiated regulatory strategies, and establish a disclosure mechanism characterised by 'professional identification and reputational penalties' to mitigate such peer AI washing behaviour.